\documentstyle[11pt,fleqn,epsf]{article}
\def\ref{par\noindent\hangindent=6mm\hangafter=1}
\begin{document}
\vbox{
\rightline{IFUG-95/07}
\rightline{cond-mat/9505051}
\rightline{Nuovo Cimento B 110, 883-885 (1995)}
}
\baselineskip 8mm
\begin{center}
{\bf BISTABILITY FOR ASYMMETRIC DISCRETE RANDOM WALKS}

\bigskip

Marco Reyes\cite{byline}

{\it Departamento de F\'{\i}sica, CINVESTAV, Apdo Postal 14-740,
M\'exico D.F., M\'exico}

H. Rosu\cite{byline}

{\it Instituto de F\'{\i}sica de la Universidad de Guanajuato, Apdo Postal
E-143, Le\'on, Gto, M\'exico}

Octavio Obreg\'on\cite{byline}

{\it Instituto de F\'{\i}sica de la Universidad de Guanajuato, Apdo Postal
E-143, Le\'on, Gto, M\'exico and Departamento de F\'{\i}sica, Universidad
Aut\'onoma Metropolitana, Iztapalapa, Apdo Postal 55-534, 09340,
Distrito Federal, M\'exico}

\end{center}

\bigskip
\bigskip

\begin{abstract}

We show that asymmetric time-continuous discrete random walks can display
bistability for
equal values of Jauslin's shifting parameters. The bistability becomes more
pronounced at increased asymmetry parameter.

\end{abstract}
\bigskip



\vskip 2cm

In a previous paper \cite{rr}, two of us presented an extension of a
supersymmetric analysis of the three-site discrete master equation due to
Jauslin \cite{j}. The essential procedure in this approach is the so-called
{\em addition of eigenvalues (shifting parameters)} method that one may
consider as a variant of
the negative factorization energy case in the Witten susy scheme as we showed
in \cite{rr}. Jauslin showed that the supersymmetric master evolution
can be put into the following iterative form

$$H_{k-1}\phi _{k-1}=-\lambda _k \phi _{k-1}, \; k\geq 1  \eqno(1) $$
This iterative procedure is based on
the initial ($k=1$) master ``Hamiltonian" operator which for homogeneous
random walks (jump rates not depending on site) is \cite{rr}
$$H_{0}= -(c_1c_2)^{1/2}\frac{\partial ^2}{\partial n^2}+
c_1+c_2 - 2(c_1c_2)^{1/2} 
     \eqno(2)  $$
$c_1$ and $c_2$ are the forward and the backward jump rates, respectively.

Already at the second step ($k=2$) Jauslin's method generates
bistability in the stationary probability ($H_2 P_2^{st}=0$) if the second
eigenvalue is chosen
to be at least two orders of magnitude smaller than the first one as shown in
\cite{j} for the case of free random walks ($c_1=c_2=1/2$)
and confirmed by us \cite{rr}.
Multistability can be produced by adding more eigenvalues.
This is of course an interesting way of generating bistability and
multistability in a master equation.

In this note we want to report on generating bistability in Jauslin's
scheme with the two eigenvalues $\lambda _1$ and $\lambda _2$ chosen to be
equal. We have found that this
is possible for asymmetric (or driven)
one-step discrete random walks, e.g., possessing jump rates of
the type
$c_1=\frac{1}{2}(1+\epsilon)$ and $c_2=\frac{1}{2}(1-\epsilon)$, with
$\epsilon$ a bias parameter.
The idea is
to see what happens if we keep the eigenvalues equal (no bistability) but
change the other
control parameter of the problem which is the degree of asymmetry (or the bias)
$\epsilon$. We have found a nice build-up of bistability with increasing
$\epsilon$ (see the figure) for the stationary state $P_2^{st}$

$$P_2^{st}(n)=const\times \frac{2}{(1-\epsilon ^2)^{1/2}
\phi _1(n)\phi _1(n-1)} \eqno(3)$$
with
$$ \phi _1(n)= \frac{(1-\epsilon ^2)^{1/4}}{\sqrt 2}
[(a(n))^{-1/2} \sinh (\gamma _2n)
- (a(n))^{1/2}\sinh(\gamma _2(n-1))]  \eqno(4)$$
where $a(n)=\frac{\cosh(\gamma _1n)}{\cosh(\gamma _1(n-1))}$, and
$\gamma _1=$ arccosh$[\frac{(1+\lambda _1)}{(1-\epsilon ^2)^{1/2}}]$,
$\gamma _2 =$ arccosh$[\frac{(1+\lambda _1 +
\lambda _2)}{(1-\epsilon ^2)^{1/2}}]$, with $\lambda _1$ and
$\lambda _2$ being Jauslin's shifting parameters.

As we said, $P_2^{st}$ is the stationary state of the $H_2$ ``Hamiltonian"
which is of the same type as $H_0$, i.e.,

$$H_{2}= -(c_1^{''}c_2^{''})^{1/2}\frac{\partial ^2}{\partial n^2}+
1 - 2(c_1^{''}c_2^{''})^{1/2}  \eqno(5) $$
with $c_1^{''}=\frac{1}{2}(1-\epsilon ^2)^{1/2}\frac{\phi _{1}(n-1)}
{\phi _{1}(n)}$ and
$c_2^{''}=\frac{1}{2}(1-\epsilon ^2)^{1/2}\frac{\phi _{1}(n)}{\phi _{1}(n-1)}$.

A possible application may be to the one-dimensional hopping propagation
of charge carriers, such as in (quasi) one-dimensional polymer chains, in
the presence
of a bias field. In this case the relation between the asymmetry parameter
and the electric field is $(1-\epsilon)/(1+\epsilon)=\exp(-eEa/kT)$,
where $E$ is
the bias field, $a$ is the lattice constant, and $T$ is the thermodynamic
temperature. One concludes that bistability can show up either in strong
fields or at low temperatures.

\section*{Acknowledgment}
This work was partially supported by the CONACyT Projects 4862-E9406 and
4868-E9406.

M.R. was supported by a CONACyT Graduate Fellowship.



%

\bigskip
{\bf Figure Caption}
\bigskip

Fig. 1  Stationary states $P_2^{st}(n)$ for an asymmetric discrete RW with
shifting parameters

$\lambda _1=\lambda _2 =0.01$
and asymmetry parameter $\epsilon$ as follows:

a) $\epsilon =0 $; b) $\epsilon =0.25$; c) $\epsilon =0.50$;
d) $\epsilon=0.75 $

\newpage
\centerline{
\epsfxsize=280pt
\epsfbox{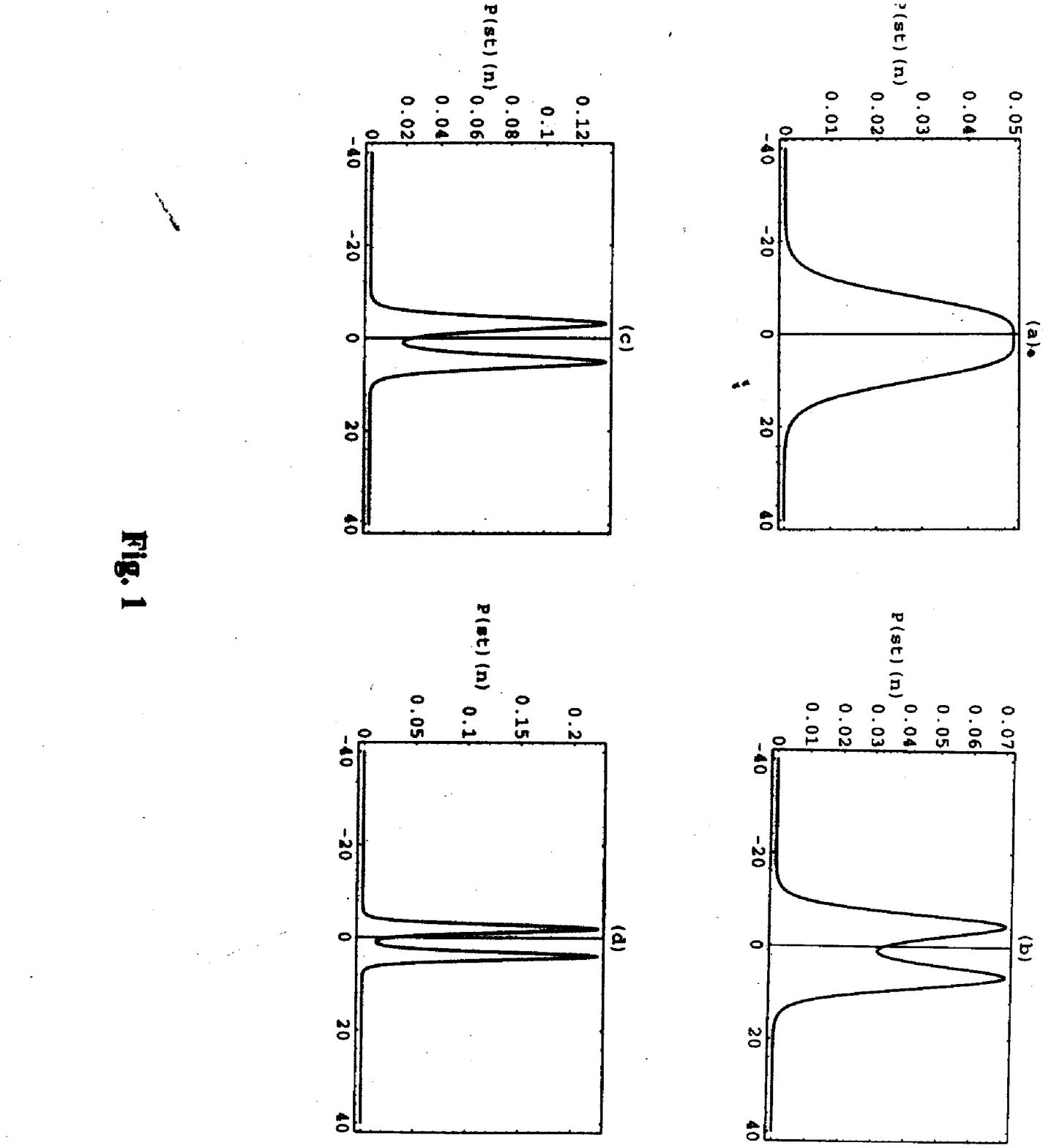}}
\vskip 4ex

\end{document}